\documentstyle[12pt]{article}

\def\GeV{\,{\rm GeV}}

\def\MeV{\,{\rm MeV}}
\def\sec{\,{\rm sec}}

\def\rcm{\,{\rm cm}}

\def\erg{{\,\rm erg}}
\def\cmm2{{\,\rm cm^{-2}}}
\def\cm2{{\,{\rm cm}^2}}
\def\cmm3{{\,{\rm cm}^{-3}}}
\def\gcmm3{{\,{\rm g\,cm^{-3}}}}

\def\mpl{{m_{\rm Pl}}}

\def\be{\begin{equation}}
\def\ee{\end{equation}}

\def\la{\mathrel{\mathpalette\fun <}}
\def\ga{\mathrel{\mathpalette\fun >}}
\def\fun#1#2{\lower3.6pt\vbox{\baselineskip0pt\lineskip.9pt
  \ialign{$\mathsurround=0pt#1\hfil##\hfil$\crcr#2\crcr\sim\crcr}}}

\begin{document}
\pagestyle{empty}
\begin{center}
\rightline{FERMILAB--Pub--93/236-A}
\rightline{astro-ph/yymmdd}
\rightline{submitted to {\it Physical Review D}}
\vspace{.2in}

{\Large \bf PRIMORDIAL NUCLEOSYNTHESIS \\
\bigskip
WITH A DECAYING TAU NEUTRINO}\\

\vspace{.2in}

Scott Dodelson,$^1$ Geza Gyuk,$^{2}$ and Michael S. Turner$^{1,2,3}$ \\
\vspace{.1in}

$^1${\it NASA/Fermilab Astrophysics Center
Fermi National Accelerator Laboratory, Batavia, IL  60510-0500}\\
\medskip
$^2${\it Department of Physics\\
The University of Chicago, Chicago, IL  60637-1433}\\
\medskip
$^3${\it Department of Astronomy \& Astrophysics\\
Enrico Fermi Institute,
The University of Chicago, Chicago, IL 60637-1433}\\
\end{center}

\vspace{.3in}

\centerline{\bf ABSTRACT}
\bigskip

\noindent  A comprehensive study of the effect of an
unstable tau neutrino on primordial nucleosynthesis is presented.
The standard code for nucleosynthesis is modified to allow for
a massive decaying tau neutrino whose daughter products include
neutrinos, photons, $e^\pm$ pairs, and/or noninteracting
(sterile) daughter products.  Tau-neutrino decays influence
primordial nucleosynthesis in three distinct ways:
(i) the energy density of the decaying tau neutrino
and its daughter products affect the expansion rate
tending to increase $^4$He, D, and $^3$He production;
(ii) electromagnetic (EM) decay products heat the EM plasma
and dilute the baryon-to-photon ratio tending to
decrease $^4$He production and increase D and
$^3$He production; and (iii) electron
neutrinos and antineutrinos produced by tau-neutrino decays
increase the weak rates that govern the neutron-to-proton
ratio, leading to decreased $^4$He production for short
lifetimes ($\la 30\sec$) and masses less than about $10\MeV$ and increased
$^4$He production for long lifetimes or large masses.
The precise effect of a decaying
tau neutrino on the yields of primordial nucleosynthesis and the
mass-lifetime limits that follow depend crucially
upon decay mode.   We identify four generic decay modes that serve to
bracket the wider range of possibilities:  tau neutrino decays to
(1) sterile daughter products (e.g., $\nu_\tau \rightarrow
\nu_\mu + \phi$; $\phi$ is a very weakly interacting
scalar particle); (2) sterile daughter product(s) + daughter
products(s) that interacts electromagnetically (e.g., $\nu_\tau
\rightarrow \nu_\mu +\gamma$); (3) electron neutrino +
sterile daughter product(s) (e.g., $\nu_\tau \rightarrow
\nu_e +\phi$); and (4) electron neutrino + daughter product(s) that
interact electromagnetically ($\nu_\tau \rightarrow \nu_e
+e^\pm$).  Mass-lifetime limits are derived for
the four generic decay modes assuming that the abundance
of the massive tau neutrino is determined by its
electroweak annihilations.  In general, nucleosynthesis excludes a tau-neutrino
of
mass $0.4\MeV -30\MeV$ for lifetimes greater than about $300\sec$.  These
nucleosynthesis bounds are timely since the current laboratory upper bounds to
the tau-neutrino mass are around $30\MeV$, and together the
two bounds very nearly exclude a long-lived
tau neutrino more massive than about $0.4\MeV$.
Further, our nucleosynthesis
bounds together with other astrophysical and laboratory bounds
exclude a tau neutrino of mass $0.4\MeV - 30\MeV$ of any lifetime
that decays with EM daughter product(s).
We use our results to constrain the mass times relic abundance
of a hypothetical, unstable species with similar
decay modes.  Finally, we note that
a tau neutrino of mass $1\MeV$ to $10\MeV$
and lifetime $0.1\sec -10\sec$ whose decay products include
an electron neutrino can reduce the
$^4$He yield to less than that for two massless neutrino species.  This
fact could be relevant if the primordial mass fraction
of $^4$He is found to be less than about 0.23 and can also lead to
a modification of the nucleosynthesis bound to the number of
light ($\ll 1\MeV$) neutrino (and other) particle species.

\newpage
\pagestyle{plain}
\setcounter{page}{1}
\section{Introduction}

Primordial nucleosynthesis is one of the cornerstones
of the hot big-bang cosmology.  The agreement between
the predictions for the abundances of D, $^3$He, $^4$He
and $^7$Li and their inferred primordial abundances provides
the big-bang cosmology's earliest, and perhaps most, stringent test.
Further, big-bang nucleosynthesis has been used to provide
the best determination of the baryon density \cite{ytsso,walker}
and to provide crucial tests of particle-physics theories, e.g.,
the stringent bound to the number of light
neutrino species \cite{nulimit,mathews}.

Over the years various aspects of
the effect of a decaying tau neutrino on primordial nucleosynthesis
have been considered \cite{ks,st1,st2,st,kaw,ketal,dol,osu}.
Each previous study focused on a specific
decay mode and incorporated different microphysics.  To be sure,
no one study was complete or exhaustive.  Our purpose here is to consider
all the effects of a decaying tau neutrino on nucleosynthesis
in an comprehensive and coherent manner.  In particular,
for the first time interactions of decay-produced electron
neutrinos and antineutrinos, which can be important for
lifetimes shorter than $100\sec$ or so, are taken into account.

The nucleosynthesis limits to the mass of an unstable tau
neutrino are currently of great interest as the best laboratory
upper mass limits \cite{labmass}, $31\MeV$ by
the ARGUS Collaboration and $32.6\MeV$ by the CLEO
Collaboration,\footnote{Both are 95\% C.L.
mass limits based upon end-point analyses of tau decays to
final states containing five pions.  The CLEO data set contains
113 such decays and the ARGUS data set contains 20 such decays \cite{labmass}.}
are tantalizingly close to the mass range excluded by nucleosynthesis,
approximately $0.4\MeV$ to $30\MeV$ for lifetimes
greater than about $300\sec$.  If the upper range
of the cosmologically excluded band can be
convincingly shown to be greater than the upper
bound to the mass from laboratory experiments, the two bounds
together imply that a long-lived tau-neutrino
must be less massive than about $0.4\MeV$.  This was the major
motivation for our study.

The effects of a massive, decaying tau neutrino on primordial
nucleosynthesis fall into
three broad categories:  (i) the energy density of the tau neutrino
and its daughter product(s) increase the expansion rate, tending
to increase $^4$He, D, and $^3$He production; (ii)  the
electromagnetic (EM) plasma is heated by the
daughter product(s) that interact electromagnetically
(photons and $e^\pm$ pairs), diluting the baryon-to-photon
ratio and decreasing $^4$He production and increasing
D and $^3$He production; and (iii) electron neutrino
(and antineutrino) daughters increase
the weak interaction rates that govern the neutron-to-proton
ratio, leading to decreased $^4$He production for short lifetimes
($\la 30\sec$) and masses less than about $10\MeV$
and increased $^4$He production for long lifetimes.
Decays that take place long after nucleosynthesis ($\tau_\nu
\sim 10^5\sec -10^6\sec$) can lead to the destruction of
the light elements through fission
reactions and additional constraints \cite{fission}, neither
of which are considered here.

In terms of the effects on primordial nucleosynthesis
there are, broadly speaking, four generic decay modes:
\begin{enumerate}

\item Tau neutrino decays to daughter products that
are all sterile, e.g., $\nu_\tau \rightarrow \nu_\mu
+\phi$ ($\phi$ is a very weakly interacting boson).
Here, only effect (i) comes into play.  Aspects of this case
were treated in Refs.~\cite{ks,st2,ketal,dol,osu}; the very recent
work in Ref.~\cite{osu} is the most complete study of this mode.

\item Tau neutrino decays to a sterile daughter product(s)
plus a daughter product(s) that interacts electromagnetically,
e.g., $\nu_\tau \rightarrow \nu_\mu + \gamma$.  Here,
effects (i) and (ii) come into play.  This case was
treated in Ref.~\cite{st1}, though not specifically
for a decaying tau neutrino.

\item Tau neutrino decays into an electron neutrino and sterile
daughter product(s), e.g., $\nu_\tau \rightarrow \nu_e
+\phi$.  Here, effects (i) and (iii) come into play.  This case
was treated in Ref.~\cite{st}; however, the interactions
of electron neutrinos and antineutrinos with the ambient
thermal plasma were not taken into account.  They can be important:
The interaction rate of a high-energy electron neutrino produced
by the decay of a massive tau neutrino relative to the
expansion rate $\Gamma /H \sim (m_\nu /\MeV)(\sec /t)$.

\item Tau neutrino decays into an electron neutrino
and daughter product(s) that interact electromagnetically,
e.g., $\nu_\tau \rightarrow \nu_e +e^\pm$.  Here,
all three effects come into play.   Aspects of this case were
treated in Ref.~\cite{kaw}, though interactions of
electron neutrinos and antineutrinos with the ambient thermal
plasma were neglected and the $\nu_e$-spectrum
was taken to be a delta function.

\end{enumerate}

{\it As we shall emphasize more than once, the effect of a tau neutrino of a
given mass and lifetime---and therefore limits to its
mass/lifetime---depends very much upon decay mode.}

\medskip

While these four generic decay modes serve to bracket
the possibilities, the situation is actually somewhat more complicated.
Muon neutrinos are not completely
sterile, as they are strongly coupled
to the electromagnetic plasma down to temperatures of order
a few MeV (times of order a fraction of a second), and thus
can transfer energy to the electromagnetic plasma.  However,
for lifetimes longer than a few seconds, their interactions with the
electromagnetic plasma are not very significant (see Ref.~\cite{sd}),
and so to a reasonable approximation muon-neutrino
daughter products can be considered sterile.
Precisely how much electromagnetic entropy is produced
and the effect of high-energy neutrinos on the proton-neutron
interconversion rates depend upon the energy distribution
of the daughter products and their interactions with the
ambient plasma (photons, $e^\pm$ pairs, and neutrinos), which
in turn depends upon the number of daughter products and
the decay matrix element.

Without going to extremes,
one can easily identify more than ten possible decay modes.
However, we believe the four generic decay modes
serve well to illustrate how the nucleosynthesis
mass-lifetime limits depend upon the decay mode and provide
reasonable estimates thereof.  In that regard,
input assumptions, e.g., the acceptable range for the
primordial abundances and the relic neutrino
abundance\footnote{The variation between different calculations
of the tau-neutrino abundance are of the order of 10\% to
20\%; they arise to different treatments of thermal averaging,
particle statistics, and so on.  Since we use
the asymptotic value of the tau-neutrino abundance
our abundances are in general smaller, making our limits
more conservative.} probably lead to comparable, if not greater,
uncertainties in the precise limits.

Finally, a brief summary of our treatment of the microphysics:
(1) The relic abundance of the tau neutrino is determined by standard
electroweak annihilations and is assumed to be frozen out
at its asymptotic value during the epoch
of nucleosynthesis, thereafter decreasing due to decays only.
Because we assume that the relic abundance of the tau
neutrino has frozen out we cannot
accurately treat the case of short lifetimes, $\tau_\nu
\la (m_\nu /\MeV )^{-2}\sec$, where inverse decays
can significantly affect the tau-neutrino abundance and that of its daughter
products \cite{inversedecay}.\footnote{For generic decay mode (1) the effect
of inverse decays for short lifetimes was considered in Ref.~\cite{osu};
it leads to additional mass constraints for short lifetimes.}
(2) Sterile daughter products, other than neutrinos, are assumed to
have a negligible pre-decay abundance (if this is not true,
the nucleosynthesis limits become even more stringent).
(3) The electromagnetic energy produced by tau-neutrino
decays is assumed to be quickly thermalized and to increase the entropy
in the electromagnetic plasma according to the first law of
thermodynamics.  (4) The perturbations to the
phase-space distributions of electron and muon neutrinos
due to tau-neutrino decays and partial coupling to the electromagnetic
plasma are computed.  (5)  The change in the weak rates that interconvert
neutrons and protons due to the distorted electron-neutrino
distribution are calculated.  (6) The total energy of the Universe includes
that of photons, $e^\pm$ pairs, neutrinos, and sterile daughter product(s).

The paper is organized as follows; in the next Section we discuss
the modifications that we have made to the standard nucleosynthesis
code.  In Section 3 we present our results, discussing how
a decaying tau neutrino affects the yields of nucleosynthesis and
deriving the mass/lifetime limits for the four generic decay
modes.  In Section 4 we discuss other astrophysical and laboratory
limits to the mass/lifetime of the tau neutrino, and finish
in Section 5 with a brief summary and concluding remarks.

\section{Modifications to the Standard Code}

In the standard treatment of nucleosynthesis \cite{kawano} it is
assumed that there are three massless neutrino species that
completely decouple from the electromagnetic plasma at a
temperature well above that of the electron mass ($T \sim
10\MeV\gg m_e$).  Thus the neutrino species do not interact with
the electromagnetic plasma and do not share in the ``heating'' of
the photons when the $e^\pm$ pairs disappear.

In order to treat the most general case of a decaying tau
neutrino we have made a number of modifications to the standard
code.  These modifications are of four kinds: (1) Change the
total energy density to account for the massive tau neutrino and
its daughter products; (2) Change the first-law of
thermodynamics for the electromagnetic plasma to account for the
injection of energy by tau decays and interactions with the
other two neutrino seas; (3) Follow the Boltzmann equations for the
phase-space distributions for electron and muon neutrinos,
accounting for their interactions with one another and the
electromagnetic plasma; (4) Modify the weak interaction rates
that interconvert neutrons and protons to take account of the
perturbations to the electron-neutrino spectrum.

These modifications required tracking five quantities as a
function of $T \equiv R^{-1}$, the
neutrino temperature in the fully decoupled limit ($R=$ the
cosmic-scale factor).  They are:  $\rho_{\nu_\tau}$,
$\rho_\phi$ (where $\phi$ is any sterile, relativistic decay
product), $T_\gamma$, and $\Delta_e$ and $\Delta_\mu$, the
perturbations to the electron-neutrino and mu-neutrino
phase-space distributions.

Our calculations were done with two separate codes. The
first code tracks $\rho_{\nu_\tau}$,
$\rho_\phi$, $T_\gamma$, $\Delta_e$, and $\Delta_\mu$
as a function of $T$, for simplicity, using Boltzmann statistics.
These five quantities were then converted to functions of the
photon temperature using the $T(T_\gamma )$
relationship calculated, and their values
were then passed to the second code, a modified version
of the standard nucleosynthesis code \cite{kawano}.\footnote{The
correct statistics for all species are of course used in the
nucleosynthesis code; the five quantities are passed
as fractional changes (to the energy density, temperature and rates)
to minimize the error made by using Boltzmann statistics in the first code.}
We now discuss in more detail the four modifications.

\subsection{Energy density}

There are four contributions to
the energy density: massive tau neutrinos, sterile decay
products, two massless neutrino species, and the EM
plasma. Let us consider each in turn.

As mentioned earlier, we fix the relic abundance of tau
neutrinos assuming that freeze out occurs before nucleosynthesis
commences ($t\ll 1\sec$).  We follow Ref.~\cite{ketal} in writing
\begin{equation}
\rho_{\nu_\tau} = r \left[ \frac{\sqrt{(3.151 T )^2 +
{m_\nu}^2}}{{3.151 T }}\right] \rho_\nu(m_\nu =0)\,\exp (-t/\tau_\nu ),
\end{equation}
where $r$ is the ratio of the number density of massive neutrinos to a massless
neutrino species, the $(3.151T)^2$ term takes account of the
kinetic energy of the neutrino, and the exponential factor
takes account of decays.  The relic abundance is taken
from Ref.~\cite{ketal}; for a Dirac neutrino it is assumed
that all four degrees are freedom are populated for masses
greater than $0.3\MeV$ (see Ref.~\cite{ketal} for further
discussion).

Note that for temperatures much
less than the mass of the tau neutrino, $\rho_{\nu_\tau}/\rho_\nu
(m_\nu =0) = rm_\nu e^{-t/\tau_\nu}/3.151T$,
which increases as the scale factor
until the tau neutrinos decay; further, $rm_\nu$ determines the
energy density contributed by massive tau neutrinos and hence
essentially all of their effects on nucleosynthesis.
The relic neutrino abundance times mass ($rm_\nu$) is shown in Fig.~1.

The energy density of the sterile decay products is slightly more
complicated.  Since the $\phi$'s are massless, their energy
density is governed by
\begin{equation}  \label{eq:phi}
\frac{d\rho_\phi}{dT} = \frac{4\rho_\phi}{T} - {f_\phi\over T}
\frac{\rho_{\nu_\tau}} {H \tau_\nu},
\end{equation}
where the first term accounts for the effect of the expansion
of the Universe and the second accounts for the energy dumped into the
sterile sector by tau-neutrino decays.  The quantity $f_\phi$
is the fraction of the tau-neutrino decay energy that goes into
sterile daughters:  for $\nu_\tau \rightarrow$ all-sterile daughter
products, $f_\phi =
1$; for $\nu_\tau \rightarrow \phi + \nu_e$ or EM, $f_\phi =0.5$;
and for all other modes $f_\phi =0$.  Eq.~(\ref{eq:phi})
was integrated numerically, and
$\rho_\phi$ was passed to the nucleosynthesis code
by means of a look-up table.

The neutrino seas were the most complicated to treat.
The contribution of the neutrino seas was divided into two parts, the
standard, unperturbed thermal contribution and the perturbation
due to the slight coupling of neutrinos to the EM plasma and
tau-neutrino decays,
\begin{equation}
\rho_\nu = \rho_{\nu 0} + \delta \rho_\nu.
\end{equation}

The thermal contribution is simply $6T^4/\pi^2$ per massless
neutrino species (two in our case).  The second term is given
as an integral over the perturbation to the neutrino phase-space
distribution,
\begin{equation}
\delta \rho_\nu = \sum_{i=e,\mu} {2\over (2\pi )^3}
\int p d^3 p \Delta_i(p,t) ,
\end{equation}
where the factor of two accounts for neutrinos and antineutrinos.

Finally, there is the energy density of the EM plasma.
Since the electromagnetic plasma is in thermal equilibrium
it only depends upon $T_\gamma$:
\begin{equation}
\rho_{\rm EM} =
\frac{6{T_\gamma}^4}{\pi^2} + \frac{2{m_e}^3 T_\gamma} {\pi^2}
\left[K_1(m_e/T_\gamma ) + \frac{3 K_2(m_e/T_\gamma )}
{m_e/T_\gamma }\right] ,
\end{equation}
where $K_1$ and $K_2$ are modified Bessel functions.
We numerically compute $T_\gamma$ as a function of $T$ by
using the first law of thermodynamics.

\subsection{First law of thermodynamics}
Energy conservation in the expanding Universe is
governed by the first law of thermodynamics,
\begin{equation}\label{eq:first}
d[\rho_{\rm TOT} R^3] = -p_{\rm TOT} dR^3,
\end{equation}
where in our case $\rho_{\rm TOT} = \rho_{\rm EM} + \rho_{\nu 0} +
\delta \rho_\nu + \rho_\phi + \rho_{\nu_\tau}$, $p_{\rm TOT} =
p_{EM} + p_{\nu 0} + \delta p_\nu + p_\phi + p_{\nu_\tau}$,
$\delta p_\nu = \delta\rho_\nu /3$, $p_\phi = \rho_\phi /3$, and
\begin{equation}
p_{\rm EM} = {2T_\gamma^4\over \pi^2} +
{2m_e^2T_\gamma^2\over \pi^2}\,K_2(m_e/T_\gamma ) .
\end{equation}
Eq.~(\ref{eq:first}) can be rewritten in a more useful form,
\begin{equation}
\frac{dT_\gamma}{dt} =
{-3H\left(\rho_{\rm TOT}+ p_{\rm TOT} -4\rho_{\nu 0} /3\right)
-d(\delta\rho_\nu + \rho_\phi + \rho_{\nu_\tau})/dt \over
d\rho_{\rm EM}/dT_\gamma}.
\end{equation}
The quantity $d\rho_{\rm EM}/dT_\gamma$ is easily calculated,
and the time derivatives of the densities can either be solved for
analytically, or taken from the previous time step.

\subsection{Neutrino phase-space distribution functions}

The Boltzmann equations governing the neutrino phase-space
distribution functions in the standard case
were derived and solved in Ref.~\cite{sd}.
We briefly summarize that treatment here, focusing on the
modifications required to include massive tau-neutrino decays.

We start with the Boltzmann equation for the phase-space
distribution of neutrino species $a$ in the absence of decays:
\begin{eqnarray}\label{eq:boltzmann}
\frac{\partial f_a}{\partial t} -{H |p|^2\over E_a}
\frac{\partial f_a}{\partial E} & = & -\frac{1}{2E_a}
\sum_{processes} \int d\Pi_1 d\Pi_2 d\Pi_3(2\pi)^4
\delta^4({p}_a+{p}_1-{p}_2-{p}_3)\nonumber\\
&\times & |{\cal M}_{a+1\leftrightarrow 2+3}|^2 [f_a f_1-f_2 f_3],
\end{eqnarray}
where the processes summed over include
all the standard electroweak $2\leftrightarrow 2$ interactions
of neutrinos with themselves and the electromagnetic plasma, and
Boltzmann statistics have been used throughout.

We write the distribution functions for the electron and
muon neutrinos as an unperturbed part plus a small perturbation:
\begin{equation}
f_i (p,t) = \exp (-p/T) + \Delta_i(p,t),
\end{equation}
where we have assumed that both species are effectively
massless.   During nucleosynthesis the photon temperature
begins to deviate from the neutrino temperature $T$, and
we define
$$\delta (t) = T_\gamma /T -1.$$

Eq.~(\ref{eq:boltzmann}) is expanded to lowest order in
$\Delta_i$ and $\delta (t)$, leading to master equations of the form:
\begin{equation}\label{eq:master}
\frac{p}{T}{\dot\Delta}_i(p,t) = 4 G_F^2 T^3
[-A_i(p,t)\Delta_i(p,t)  + B_i(p,t)\delta(t)
 +C_i(p,t) + C_i^\prime(p,t)],
\end{equation}
where $i=e,\mu$ and the expressions for $A_i$, $B_i$, $C_i$,
and $C_i^\prime$ are given in Ref.~\cite{sd} [in Eq.~(2.11d)
for $C_\mu$ the coefficient $(c+8)$ should be $(c+7)$].

In context of tau-neutrino decays we treat decay-produced
muon neutrinos as a sterile species, and thus
we are only interested in modifying the master equation for electron
neutrinos to allow for decays.  In the case of two-body decays
(e.g., $\nu_\tau \rightarrow \nu_e+\phi$
or $\nu_\tau \rightarrow \nu_e +$ EM) the additional term
that arises on the right-hand side of Eq.~(\ref{eq:master}) is
\begin{equation}
{p\over T}{\dot\Delta}_e(p,T) = \cdots +
{n_{\nu_\tau}\over \tau_\nu}\,{2\pi^2 \over pT}\,\delta (p-m_\nu /2),
\end{equation}
where $n_{\nu_\tau}$ is the number density of massive tau
neutrinos.\footnote{For $m_\nu$ we actually use our expression
for the total tau-neutrino energy $E_\nu= \sqrt{m_\nu^2+(3.151T)^2}$.
Except for very short lifetimes and small masses, $E_\nu\approx m_\nu$.}

The decay mode $\nu_\tau \rightarrow \nu_e +e^\pm$ has a three-body
final state, so that the energy distribution of electron neutrinos
is no longer a delta function.  In this case, the source term is
\begin{equation}
{p\over T}{\dot\Delta}_e(p,T) = \cdots +
{32\pi^2 n_{\nu_\tau} p (3-4p/m_\nu ) \over \tau_\nu m_\nu^3 T}
\,\theta (p-m_\nu /2),
\end{equation}
where for simplicity we have assumed that all particles except
the massive tau neutrino are ultrarelativistic.

\subsection{Weak-interaction rates}

Given $\Delta_e $, it is simple to
calculate the perturbations to the weak interaction rates
that convert protons to neutrons and vice versa
(see Ref.~\cite{sd} for details). The perturbations to
the weak rates are obtained by substituting $\exp (-p/T) +\Delta_e (p,t)$
for the electron phase-space distribution in the usual expressions
for the rates \cite{sd} and then expanding to lowest order.
The perturbations to the rates for proton-to-neutron conversion
and neutron-to-proton conversion (per nucleon) are respectively
\begin{eqnarray}
\delta\lambda_{pn} & = & \frac{1}{\lambda_0 \tau_n} \int_{m_e}^\infty EdE
(E^2-m_e^2)^{1/2} (E + Q)^2 \Delta_e(E+Q), \\
\delta\lambda_{np} & = & \frac{1}{\lambda_0 \tau_n} \int_{m_e}^\infty EdE
(E^2-m_e^2)^{1/2} (E - Q)^2 \Delta_e(E-Q),
\end{eqnarray}
where Boltzmann statistics have been used for all species,
$\tau_n$ is the neutron mean lifetime, $Q=1.293\MeV$ is the
neutron-proton mass difference, and
$$\lambda_0 \equiv \int_{m_e}^Q EdE(E^2-m_e^2)^{1/2}(E-Q)^2 .$$

The perturbations to the weak rates are computed in the first
code and passed to the nucleosynthesis code by means of a
look-up table.  The unperturbed part of the weak rates are
computed by numerical integration in the nucleosynthesis
code; for all calculations we took the neutron mean lifetime to be $889\sec$.

\section{Results}

In this section we present our results for the four generic decay modes.
Mode by mode we discuss how the light-element abundances
depend upon the mass and lifetime of the tau neutrino
and derive mass/lifetime limits.  We exclude a mass and
lifetime if, for no value of the baryon-to-photon ratio,
the light-element abundances can satisfy:
\begin{eqnarray}
Y_P & \le & 0.24 ;\\
{\rm D/H} & \ge & 10^{-5}; \\
({\rm D} +^3{\rm He})/{\rm H} & \le & 10^{-4};\\
{\rm Li}/{\rm H} &\le & 1.4\times 10^{-10}.
\end{eqnarray}
For further discussion of this choice of constraints to
the light-element abundances we refer the reader
to Ref.~\cite{walker}.

The $^4$He and D + $^3$He abundances play the most
important role in determining the excluded regions.
The mass/lifetime limits that follow necessarily
depend upon the range of acceptable primordial abundances that one
adopts, a fact that should be kept in mind when comparing
the work of different authors and assessing confidence levels.
Further, the relic abundances used by different
authors differ by 10\% to 20\%.
Lastly, the precise limit for a specific decay mode
will of course differ slightly from that derived for its ``generic class.''

In illustrating how the effects of a decaying tau neutrino
depend upon lifetime and in comparing different decay modes we
use as a standard case an initial (i.e., before decay
and $e^\pm$ annihilations) baryon-to-photon ratio $\eta_i =
8.25\times 10^{-10}$.  In the absence of entropy production
(no decaying tau neutrino or decay modes 1 and 3 which produce
no EM entropy) the final
baryon-to-photon ratio $\eta_0 = 4\eta_i /11 = 3\times 10^{-10}$,
where $4/11$ is the usual factor that arises due to the entropy transfer
from $e^\pm$ pairs to photons.  In the case of decay modes
2 and 4 there can be significant EM entropy production, and
the final baryon-to-photon ratio $\eta = \eta_0/(S_f/S_i)
\le \eta_0$ ($S_f/S_i$ is the ratio of the EM entropy per
comoving volume after decays to that before decays).
Even though $\eta_0$ does not
correspond to the present baryon-to-photon ratio if there has
been entropy production, we believe
that comparisons for fixed $\eta_0$ are best for
isolating the three different effects of a decaying tau
neutrino on nucleosynthesis.  For reference,
in the absence of a decaying tau neutrino the $^4$He mass
fraction for our standard case is:  $Y_P=0.2228$ (two massless
neutrino species) and $0.2371$ (three massless neutrino species).

\subsection{$\nu_\tau \rightarrow$ sterile daughter products}

Since we are considering lifetimes greater than $0.1\sec$, by which time muon
neutrinos are essentially decoupled, the muon neutrino is by our definition
effectively sterile, and examples of this decay mode include,
$\nu_\tau \rightarrow \nu_\mu + \phi$ where $\phi$ is some
very weakly interacting scalar particle (e.g., majoron)
or $\nu_\tau \rightarrow \nu_\mu +\nu_\mu + {\bar\nu}_\mu$.

For this decay mode the only effect of the unstable tau
neutrino on nucleosynthesis involves the energy density it and
its daughter products contribute.  Thus, it is the simplest
case, and we use it as ``benchmark'' for comparison to the
other decay modes.  The light-element abundances
as a function of tau-neutrino lifetime are shown in Figs.~2-4
for a Dirac neutrino of mass $20\MeV$.

The energy density of the massive
tau neutrino grows relative to a massless neutrino species
as $rm_\nu/3T$ until the tau neutrino decays, after
which the ratio of energy density in the daughter products
to a massless neutrino species remains constant.
For tau-neutrino masses in the $0.3\MeV$ to $30\MeV$ mass
range and lifetimes greater than about a second the energy
density of the massive tau neutrino exceeds that of a massless neutrino
species before it decays, in spite of its smaller abundance
(i.e., $r\ll 1$). The higher energy density increases the
expansion rate and ultimately $^4$He production because
it causes the neutron-to-proton ratio to freeze out earlier and at
a higher value and because fewer neutrons decay before nucleosynthesis
begins.  Since the neutron-to-proton ratio freezes out around
$1\sec$ and nucleosynthesis occurs at
around a few hundred seconds, the $^4$He abundance is only sensitive
to the expansion rate between one and a few hundred seconds.

In Fig.~2 we see that for short lifetimes ($\tau_\nu\ll 1\sec$)
the $^4$He mass fraction approaches that for two massless neutrinos
(tau neutrinos decay before their energy density becomes
significant).  As expected, the $^4$He mass fraction increases with
lifetime leveling off at a few hundred
seconds at a value that is significantly greater than that
for three massless neutrino species.

The yields of D and $^3$He depend upon how much of these isotopes
are not burnt to $^4$He.  This in turn depends upon competition
between the expansion rate and nuclear reaction rates:  Faster expansion
results in more unburnt D and $^3$He.  Thus the yields of D and
$^3$He increase with tau-neutrino lifetime, and begin to level
off for lifetimes of a few hundred seconds as this is when
nucleosynthesis is taking place (see Fig.~3).

The effect on the yield of $^7$Li is a bit more complicated.
Lithium production decreases with increasing $\eta$
for $\eta \la 3\times 10^{-10}$ because the final abundance
is determined by competition between the expansion rate and
nuclear processes that destroy $^7$Li, and increases
with increasing $\eta$ for $\eta \ga 3\times 10^{-10}$
because the final abundance is determined by competition between
the expansion rate and nuclear processes that produce $^7$Li.
Thus, an increase in expansion rate leads to increased $^7$Li
production for $\eta \la 3\times 10^{-10}$ and decreased $^7$Li
production for $\eta \ga 3\times 10^{-10}$; this is shown
in Fig.~4.  Put another way
the valley in the $^7$Li production curve shifts to
larger $\eta$ with increasing tau-neutrino lifetime.

We show in Figs.~5 and 6 the excluded
region of the mass/lifetime plane for a Dirac
and Majorana tau neutrino respectively.
As expected, the excluded mass range grows with lifetime,
asymptotically approaching $0.3\MeV$ to
$33\MeV$ (Dirac) and $0.4\MeV$ to
$30\MeV$ (Majorana).  We note the significant dependence of the excluded
region on lifetime; our results are in good agreement with the
one other work where comparison is straightforward \cite{ketal},
and in general agreement with Refs.~\cite{st2,osu}.

\subsection{$\nu_\tau \rightarrow$ sterile + electromagnetic daughter products}

Again, based upon our definition of sterility, the sterile
daughter could be a muon neutrino; thus, examples of this
generic decay mode include $\nu_\tau \rightarrow
\nu_\mu + \gamma$ or $\nu_\tau \rightarrow \nu_\mu + e^\pm$.
Our results here are based upon a two-body decay (e.g.,
$\nu_\tau \rightarrow \nu_\mu + \gamma$), and
change only slightly in the case of a three-body decay
(e.g., $\nu_\tau \rightarrow \nu_\mu + e^\pm$), where a larger
fraction of the tau-neutrino mass goes into electromagnetic entropy.

Two effects now come into play:  the energy density
of the massive tau neutrino and its daughter products speed
up the expansion rate, tending to increase $^4$He, $^3$He, and D
production;  and EM entropy production due to tau-neutrino decays reduce
the baryon-to-photon ratio (at the time of nucleosynthesis),
tending to decrease $^4$He production
and to increase D and $^3$He production.  Both
effects tend to shift the $^7$Li valley (as a function
of $\eta_0$) to larger $\eta_0$.

While the two effects have the ``same sign'' for D, $^3$He, and
$^7$Li, they have opposite signs for $^4$He.
It is instructive to compare $^4$He production as a function of
lifetime to the previous ``all-sterile'' decay mode.  Because of
the effect of entropy production, there is little
increase in $^4$He production
until a lifetime greater than $1000\sec$ or so.  For
lifetimes greater than $1000\sec$ the bulk of the entropy
release takes place after nucleosynthesis,
and therefore does not affect the value of $\eta$ during nucleosynthesis.

Because of the competing effects on $^4$He production,
the impact of an unstable,
massive tau neutrino on nucleosynthesis is significantly less
than that in the all-sterile decay mode for lifetimes less than
about $1000\sec$.  The excluded region of the mass/lifetime
plane is shown in Figs.~5 and 6.  For lifetimes greater than about $1000\sec$
the excluded mass interval is essentially the same as
that for the all-sterile decay mode; for shorter lifetimes it
is significantly smaller.

Finally, because of entropy production, the final value of the
baryon-to-photon ratio is smaller for fixed initial
baryon-to-photon ratio:  it is reduced by the
factor by which the entropy per comoving volume is increased.
In the limit of significant entropy production ($S_f/S_i
\gg 1$), this factor is given by, cf. Eq. (5.73) of Ref.~\cite{kt},
\begin{equation}\label{eq:entropy}
S_f/S_i \simeq 0.13 rm_\nu \sqrt{\tau_\nu /\mpl} \simeq 1.5
\,{rm_\nu \over \MeV}\,\sqrt{\tau_\nu \over 1000 \sec}.
\end{equation}
A precise calculation of entropy production for this decay
mode is shown in Fig.~7.  As can be seen in the figure
or from Eq.~(\ref{eq:entropy}), entropy production becomes
significant for lifetimes longer than about $100\sec$.

\subsection{$\nu_\tau \rightarrow \nu_e$ + sterile daughter products}

Once again, by our definition of sterility this includes
decay modes such as $\nu_\tau \rightarrow \nu_e + \phi$
or $\nu_\tau \rightarrow \nu_e +\nu_\mu {\bar\nu}_\mu$.
Here, we specifically considered the two-body decay mode $\nu_\tau
\rightarrow \nu_e + \phi$, though the results for the
three-body mode are very similar.

Two effects come into play:  the energy density
of the massive tau neutrino and its daughter products
and the interaction of daughter electron
neutrinos with the nucleons and the ambient plasma.
The first effect has been discussed previously.
The second effect leads to some interesting new effects.

Electron neutrinos and antineutrinos produced by tau-neutrino decays
increase the weak rates that govern the neutron-to-proton
ratio.  For short lifetimes ($\la 30\sec$) and masses less
than about $10\MeV$ the main effect is to delay slightly
the ``freeze out'' of the neutron-to-proton ratio, thereby
decreasing the neutron fraction at the time of nucleosynthesis
and ultimately $^4$He production.  For long lifetimes, or short lifetimes
and large masses, the perturbations to the $n\rightarrow p$
and $p\rightarrow n$ rates (per nucleon) are comparable; since after freeze out
of the neutron-to-proton ratio there are about six times as
many protons as neutrons, this has the effect of increasing
the neutron fraction and $^4$He production.  This is illustrated in Fig.~8.
The slight shift in the neutron fraction does not affect
the other light-element abundances significantly.

The excluded portion of the mass/lifetime plane is shown
in Figs.~5 and 6.  It agrees qualitatively with the results
of Ref.~\cite{st}.\footnote{The authors of Ref.~\cite{st}
use a less stringent constraint to $^4$He production,
$Y_P\le 0.26$; in spite of this, in some regions of
the $m_\nu -\tau_\nu$ plane their bounds are as, or even more, stringent.
This is presumably due to the neglect of electron-neutrino
interactions with the ambient plasma.}  Comparing the limits for this
decay mode with the all-sterile mode, the effects of electron-neutrino
daughter products are clear:  for long lifetimes
much higher mass tau neutrinos are excluded
and for short lifetimes low-mass tau neutrinos are allowed.

\subsection{$\nu_\tau \rightarrow \nu_e$ + electromagnetic daughter products}

Now we consider the most complex of the decay modes, where
none of the daughter products is sterile.
Specifically, we consider the decay mode $\nu_\tau \rightarrow
\nu_e+e^\pm$, though our results change very little for
the two-body decay $\nu_\tau \rightarrow \nu_e + \gamma$.

In this case all three effects previously discussed come into
play:  the energy density of the massive tau neutrino and
its daughter products speed up the expansion rate; the
entropy released dilutes the baryon-to-photon ratio;
and daughter electron neutrinos increase the weak-interaction
rates that control the neutron fraction.  The net effect
on $^4$He production is shown in Fig.~9 for a variety of
tau-neutrino masses.  The main difference between this
decay mode and the previous one, $\nu_\tau \rightarrow \nu_e$ +
sterile, is for lifetimes between
$30\sec$ and $300\sec$, where the increase in $^4$He production is
less due to the entropy production which
reduces the baryon-to-photon ratio at the time of nucleosynthesis.

The excluded region of the mass/lifetime plane is shown in
Figs.~5 and 6.  It agrees qualitatively with the results
of Ref.~\cite{kaw}.\footnote{The authors of Ref.~\cite{kaw}
use a less stringent constraint to $^4$He production,
$Y_P\le 0.26$; in spite of this, in some regions of
the $m_\nu -\tau_\nu$ plane their bounds are as, or even more, stringent.
This is presumably due to the neglect of electron-neutrino
interactions with the ambient plasma.}
The excluded region for this decay mode
is similar to that of the previous
decay mode, except that lifetimes less than about $100\sec$ are
not excluded as entropy production has diminished $^4$He
production in this lifetime interval.

\subsection{Limits to a generic light species}

We can apply the arguments for the four decay
modes discussed above to a hypothetical species whose
relic abundance has frozen out at a value $r$ relative
to a massless neutrino species before the epoch of
primordial nucleosynthesis (also see Refs.~\cite{st1,st2}).
The previous limits become limits to $rm$ as a function
of lifetime $\tau$ and mass $m$, which are difficult
to display.  With the exception of the effect that involves
daughter electron neutrinos, all other effects only depend
upon $rm$, which sets the energy density of the massive
particle and its daughter products.  In Fig.~10, we show
that for lifetimes greater than about $100\sec$ and masses
greater than about $10\MeV$, the $^4$He production is
relatively insensitive to the mass of the decaying particle.
This means that for lifetimes greater than about $100\sec$
the limit to $rm$ should be relatively insensitive to
particle mass.

We show in Fig.~11 the excluded regions
of the $rm$-$\tau$ plane for a $20\MeV$ decaying particle.
In deriving these limits we used the same criteria for
acceptable light-element abundances and assumed three massless
neutrino species.  The limits to $rm$ for decay modes without
electron-neutrino daughter products are strictly independent
of mass; the two other should be relatively insensitive
to the particle mass for $\tau \ga 100\sec$ (and the actual
limits are more stringent for $m > 20\MeV$).

\section{Laboratory and Other Limits}

There are a host of other constraints to the mass and lifetime
of the tau neutrino~\cite{sarkar}.
As a general rule, cosmological arguments, such as the one
presented above, pose {\it upper} limits to the tau-neutrino
lifetime for a given mass:   cosmology
has nothing to say about a particle
that decays very early since it would not have affected
the ``known cosmological history.''  Laboratory experiments
on the other hand pose {\it lower}
limits to the lifetime because nothing happens inside
a detector if the lifetime of the decaying particle is too long.
Finally, astrophysical considerations generally rule out
bands of lifetime since  ``signals'' can only be detected if (a) the tau
neutrinos escape the object of interest before decaying
and (b) decay before they pass by earthly detectors.

\subsection{Laboratory}

The most important limits of course are the direct limits to the
tau-neutrino mass. These have come down steadily over the past few
years. The current upper limits are $31\MeV$ and
$32.6\MeV$ \cite{labmass}.

If the tau neutrino has a mass greater than $2m_e = 1.02\MeV$,
then the decay $\nu_\tau
\rightarrow \nu_e+e^\pm$ takes place through ordinary
electroweak interactions at a rate
\be\label{UET}
\Gamma = {G_{F}^2 m_\nu^5\over 192\pi^3} \vert U_{e\tau} \vert^2
        \vert U_{ee} \vert^2  \simeq
{ (m_\nu /{\MeV})^5 \vert U_{e\tau} \vert^2
                \over 2.9\times 10^4 ~{\rm sec}} ,
\ee
where $U_{e\tau}$ and $U_{ee}$ are elements of the
unitary matrix that relates
mass eigenstates to weak eigenstates, the leptonic
equivalent of the Cabbibo-Kobayashi-Maskawa matrix.  We note that
the rate could be larger (or even perhaps smaller) in models where
the decay proceeds through new interactions.  Thus, limits to
$U_{e\tau}$ give rise to model-dependent limits to the tau-neutrino lifetime.

A number of experiments have set limits to $U_{e\tau}$.
The most sensitive experiment in the mass range $1.5\MeV
< m_\nu < 4\MeV$ was performed at the power reactor in Gosgen,
Switzerland~\cite{gosgen}, which produces tau
neutrinos at a rate proportional to $\vert U_{e\tau}\vert^2$
through decay of heavy nuclei and $\nu_e-\nu_\tau$ mixing.
Above this mass range, experiments that search for
additional peaks in the positron spectrum of the
$\pi^+ \rightarrow e^+\nu$ decay (due to $\nu_e-\nu_\tau$ mixing)
provide the strictest limits. In the
mass range $4\MeV < m_\nu < 20\MeV$,
Bryman et al. \cite{bryman}\
set the limits shown in Fig.~12; for larger masses the best limits
come from Ref.~\cite{leener}.

There are also direct accelerator bounds to the lifetime of a
an unstable tau neutrino that produces a photon or $e^\pm$ pair.
In particular, as has been recently emphasized by
Babu et al. \cite{babu}, the BEBC beam dump experiment~\cite{bebc}
provides model-independent limits based upon the direct search for the
EM decay products.  These limits, while not quite as
strict as those mentioned above, are of interest
since they apply to the photon mode and to the $e^\pm$ mode
even if the decay proceeds through new interactions.  The limit,
\be
\tau_\nu > 0.18\,(m_\nu /\MeV)\,\sec ,
\ee is shown in Fig.~12.

\subsection{Astrophysical}

The standard picture of type II supernovae has the binding
energy of the newly born neutron star (about $3\times
10^{53}\erg$) shared equally by neutrinos
of all species emitted from a neutrinosphere of temperature of about $4\MeV$.
There are two types of limits based upon SN 1987A, and combined they
rule out a large region of $m_\nu - \tau_\nu$ plane.

First, if the tau neutrino decayed
after it left the progenitor supergiant, which has a radius
$R\simeq 3\times10^{12}\rcm$, the high-energy daughter
photons could have been detected \cite{smm,ktsn,pvo}. The Solar Maximum
Mission (SMM) Gamma-ray Spectrometer set an upper
limit to the fluence of $\gamma$ rays during the ten
seconds in which neutrinos were detected:
\be
 f_\gamma < 0.9~{\rcm}^{-2}; \qquad 4.1\MeV < E_\gamma < 6.4\MeV.
\ee
As we will see shortly, if only one in $10^{10}$
of the tau neutrinos leaving the supernova produced a photon,
this limit would have been saturated. In the mass regime
of interest there are two ways out of
this constraint:  The lifetime can be so
long that the arrival time was more than ten seconds
after the electron antineutrinos arrived, or
the lifetime can be so short that
the daughter photons were produced
inside the progenitor.  We can take account of both of these
possibilities in the following formula for the
expected fluence of $\gamma$ rays:
\be
f_{\gamma,10} = f_{\nu\bar\nu} W_\gamma B_\gamma\langle F_1 F_2 \rangle
\ee
where the subscript $10$ reminds us that we are
only interested in the first ten seconds,
$f_{\nu\bar\nu} \simeq 1.4\times 10^{10}$ cm$^{-2}$ is the fluence of
a massless neutrino species, $W_\gamma \sim 1/4$ is the
fraction of decay photons produced with energies between
$4.1\MeV$ and $6.4\MeV$, $F_1$ is
the fraction of tau neutrinos that decay outside the progenitor,
and $F_2$ is the fraction of these that decay early enough so that the decay
products were delayed by less than ten seconds.  The quantity
$B_\gamma$ is the branching ratio to a decay mode that includes a photon.
For $m_\nu \ga 1\MeV$ one expects the $\nu_e+e^\pm$ mode to be dominant;
however, ordinary radiative corrections should lead to
$B_\gamma \simeq 10^{-3}$ \cite{mohapatra}.  Finally angular brackets denote
an average over the Fermi-Dirac distribution of neutrino momenta,
\be
\langle A\, \rangle \equiv {1\over 1.5\zeta(3) T^3}
\int_0^\infty {A\,dp\,p^2\over e^{E/T} + 1},
\ee
where $T\simeq 4$ MeV is the temperature of the neutrinosphere and
$E = (p^2 + m_\nu^2)^{1/2}$.

To evaluate the fluence of gamma rays we need to know
$F_1$ and $F_2$. The fraction $F_1$ that decay outside the
progenitor is simply $e^{-t_1/\tau_L}$ where $t_1 = R/v = RE/p$
and the ``lab'' lifetime $\tau_L
= \tau E/m_\nu$. Of these, the fraction
whose decay products arrive {\it after}
ten seconds is $e^{-t_2/\tau_L}/e^{-t_1/\tau_L}$ where $t_2 = 10\sec
/(1-v/c)$; thus, $F_2 = 1 - e^{(t_1-t_2)/\tau_L}$.
Figure 12 shows this constraint assuming
a branching ratio $B_\gamma=10^{-3}$.

The second constraint comes from observing that if tau
neutrinos decayed within the progenitor supergiant,
the energy deposited (up to about $10^{53}\erg$) would have
``heated up'' the progenitor so much as to conflict
with the observed optical luminosity of SN 1987A (and other
type II supernovae) \cite{mohapatra,schramm}.
We require
\be
E_{\rm input} = \langle (1-F_1) \rangle E_\nu \la 10^{47} \erg ,
\ee
where $E_\nu \sim 10^{53}\erg$ is the energy carried off
by a massless neutrino species, and $1-F_1$ is the fraction
of tau neutrinos that decay within
the progenitor.  This constraint is mode-independent since decay-produced
photons or $e^\pm$ pairs will equally well ``overheat'' the progenitor.
As Fig.~12 shows, the ``supernova-light'' bound is extremely powerful.

Finally, a note regarding our SN 1987A constraints.
We have assumed that a massive tau-neutrino
species has a Fermi-Dirac distribution with the same
temperature as a massless ($m_\nu \ll 10\MeV$)
neutrino species. This is almost certainly false.
Massive ($m_\nu \ga 10\MeV$ or so)
tau neutrinos will drop out of chemical equilibrium
(maintained by pair creation/annihilations and possibly
decays/inverse decays) interior to the usual neutrinosphere
as the Boltzmann factor suppresses annihilation
and pair creation rates relative to scattering rates.
This leads us to believe that we have actually {\it underestimated}
the fluence of massive neutrinos.
While the problem has yet to be treated rigorously,
we are confident that, if anything, our simplified treatment
results in limits that are overly conservative.
Accurate limits await a more detailed analysis \cite{sigl}.

\subsection{Cosmological}

The most stringent cosmological constraint for masses
$0.1\MeV \la m \la 100\MeV$ is the nucleosynthesis bound
discussed in this paper. Nonetheless, it is worthwhile to
mention some of the other cosmological limits since they are
based upon independent arguments.   A stable tau neutrino
with mass in the MeV range contributes much more energy density than is
consistent with the age of the Universe.
Such a neutrino must be unstable, with a lifetime short
enough for its decay products to lose enough most of their energy
to ``red shifting '' \cite{dicus}.
The lifetime limit is mass dependent; a neutrino with a mass of about
$1\MeV$ must have a lifetime shorter than about $10^9\sec$,
and the constraint gets less severe for larger or smaller masses.
There is an even more stringent bound based the necessity
of the Universe being matter dominated by a red shift of
about $10^4$ in order to produce the observed large-scale structure
\cite{steigman}.  Finally, there are other
nucleosynthesis bounds based upon the dissociation of the light
elements by decay-produced photons or electron-neutrinos \cite{fission}
and by $e^\pm$ pairs produced by the continuing annihilations
of tau neutrinos \cite{josh}.

\section{Summary and Discussion}

We have presented a comprehensive study of the effect of
an unstable tau neutrino on primordial nucleosynthesis.
The effects on the primordial abundances and the mass/lifetime
limits that follow depend crucially upon the decay
mode.  In the context of primordial nucleosynthesis
we have identified four generic decay modes that bracket
the larger range of possibilities:  (1)
all-sterile daughter products; (2) sterile daughter product(s)
+ EM daughter product(s); (3) $\nu_e$ + sterile daughter product(s);
and (4) $\nu_e$ + EM daughter product(s).  The excluded
regions of the tau-neutrino mass/lifetime plane for these
four decay modes are shown in Figs.~5 (Dirac) and 6 (Majorana).

In the limit of long lifetime ($\tau_\nu \gg 100\sec$), the
excluded mass range is:  $0.3\MeV -33\MeV$ (Dirac) and
$0.4\MeV - 30\MeV$ (Majorana).  Together with current
laboratory upper mass limits, $31\MeV$ (ARGUS) and $32.6\MeV$
(CLEO), our results very nearly exclude a long-lived, tau neutrino
more massive than about $0.4\MeV$.  Moreover, other
astrophysical and laboratory data exclude a tau-neutrino
in the $0.3\MeV - 50\MeV$ mass range if its decay product(s)
include a photon or $e^\pm$ pair.  Thus, if the mass of the
tau neutrino is the range $0.4\MeV$ to $30\MeV$, then its decay
products cannot include a photon or an $e^\pm$ pair and its
lifetime must be shorter than a few hundred seconds.

We note that the results of Ref.~\cite{osu} for the all-sterile
decay mode are more restrictive than ours, excluding masses
from about $0.1\MeV$ to about $50\MeV$ for $\tau_\nu \gg
100\sec$.  This traces in
almost equal parts to (i) small ($\Delta Y \simeq +0.003$), but significant,
corrections to the $^4$He mass fraction and
(ii) slightly larger relic neutrino abundance.
With regard to the first difference, this illustrates the
sensitivity to the third significant figure of the $^4$He
mass fraction.  With regard to the second difference, it is
probably correct that within the assumptions made
the tau-neutrino abundance during nucleosynthesis is
larger than what we used.  However, other effects that have
been neglected probably lead to differences in
the tau-neutrino abundance of the
same magnitude.  For example, for tau-neutrino masses
around the upper range of excluded masses, $50\MeV -100\MeV$,
finite-temperature corrections, hadronic final states
(e.g., a single pion), and tau-neutrino mixing have not
been included in the annihilation cross section and
are likely to be important at the 10\% level.

So is a tau neutrino with lifetime greater than
a few hundred seconds and mass greater than a fraction
of an $\MeV$ ruled out or not?  Unlike a limit based upon
a laboratory experiment, it is impossible to place
standard error flags on an astrophysical or cosmological
bound.  This is because of assumptions that
must be made and modeling that must be done.  For example,
the precise limits that one derives depend
upon the adopted range of acceptable light-element abundances.
To be specific, in Ref.~\cite{osu} the upper limit of the
excluded mass range drops to around $38\MeV$ and the lower
limit increases to about $0.4\MeV$ when the
primordial $^4$He mass fraction is allowed to be as large as 0.245
(rather than 0.240).  {\it In our opinion, a very strong case has been
made against a tau-neutrino mass in the mass range $0.4\MeV$ to $30\MeV$
with lifetime much greater than $100\sec$; together
with the laboratory limits this very nearly excludes a
long-lived, tau neutrino of mass greater than $0.4\MeV$.}

Perhaps the most interesting thing found in our study is the fact that
a tau neutrino of mass $1\MeV$ to $10\MeV$ and lifetime
$0.1\sec$ to $10\sec$ that decays to an electron neutrino
and a sterile daughter product can very significantly decrease
the $^4$He mass fraction (to as low as 0.18 or so).  It has
long been realized that the standard picture of nucleosynthesis
would be in trouble if the primordial $^4$He mass fraction
were found to be smaller than about 0.23; within the standard
framework we have found one way out:  an unstable tau
neutrino.\footnote{Based upon dimensional considerations the
lifetime for the mode $\nu_\tau \rightarrow \nu_e+\phi$
is expected to be $\tau_\nu \sim 8\pi f^2/m_\nu^3$,
where $f$ is the energy scale of the superweak interactions
that mediate the decay.  For $\tau_\nu \sim 10\sec$ and
$m_\nu\sim 10\MeV$, $f\sim 10^9\GeV$.}
In principle, the possibility of an unstable tau neutrino
also loosens the primordial-nucleosynthesis
bound to the number of light species which is largely
based on the overproduction of $^4$He.  However, an
unstable tau neutrino does not directly affect the primordial
important nucleosynthesis bound to the
baryon-to-photon ratio (and $\Omega_B$) as this bound involves
the abundances of D, $^3$He, and $^7$Li and not $^4$He.

Finally, we translated our results for the tau neutrino into limits to the
relic abundance of an unstable, hypothetical particle species that
decays into one of the four generic decay models discussed.
Those very stringent limits are shown in Fig.~11.

\vskip 1.5cm
\noindent  We thank Robert Scherrer, David Schramm, Gary Steigman,
and Terry Walker for useful comments.  This work was supported in part by the
DOE (at Chicago and Fermilab), by the NASA through
NAGW-2381 (at Fermilab), and GG's NSF predoctoral fellowship.
MST thanks the Aspen Center for Physics for its hospitality
where some of this work was carried out.

\newpage

\section{Figure Captions}
\medskip

\noindent {\bf Figure 1:}  The relic neutrino abundance
used in our calculations as a function of neutrino mass:
Dirac neutrino, including both helicity states
(solid curve), and Majorana neutrino (broken curve); results
from Ref.~\cite{ketal}.

\bigskip
\noindent {\bf Figure 2:}  The $^4$He yield as a function
of tau-neutrino lifetime for the four generic decay modes, a
$20\MeV$ Dirac neutrino, and a baryon-to-photon ratio that
in the absence of entropy production leads to a present
value $\eta_0 = 3\times 10^{-10}$.
For reference, the $^4$He yield in the absence of a decaying
tau neutrino is $Y_P=0.2228$ (2 massless neutrinos) and
$0.2371$ (3 massless neutrinos).

\bigskip
\noindent{\bf Figure 3:}  The D (solid) and D + $^3$He (broken)
yields as a function of tau-neutrino lifetime for the all-sterile decay mode,
$\eta = 3 \times 10^{-10}$, and a $20\MeV$ Dirac neutrino.

\bigskip
\noindent{\bf Figure 4:}  The $^7$Li yield as a function
of tau-neutrino lifetime for the all-sterile decay mode,
$\eta = 10^{-10}$ (solid) and $10^{-9}$ (broken),
and a $20\MeV$ Dirac neutrino.

\bigskip
\noindent{\bf Figure 5:}  Excluded regions of the mass-lifetime
for a Dirac neutrino and the four generic decay modes.
The excluded regions are to the right of the curves;
our results are not applicable to the region labeled N/A
as tau-neutrino inverse decays can be important and
have not been included (see Ref.~\cite{osu}).

\bigskip
\noindent{\bf Figure 6:}  Excluded regions of the mass-lifetime
for a Majorana neutrino and the four generic decay modes.
The excluded regions are to the right of the curves;
our results are not applicable to the region labeled N/A
as inverse decays can be important.

\bigskip
\noindent{\bf Figure 7:}  Entropy production as a function
of tau-neutrino lifetime for a Dirac neutrino of mass
1, 5, 10, $20\MeV$ and the $\nu_\tau \rightarrow \phi$
+ EM decay mode.  $S_f/S_i$ is the ratio of the entropy per
comoving volume before and after tau-neutrino decays.

\bigskip
\noindent{\bf Figure 8:} $^4$He yield as a function of
tau-neutrino lifetime for the $\nu_\tau \rightarrow
\nu_e + \phi$ decay mode, $\eta = 3\times 10^{-10}$, and
Dirac masses of 1, 5, 10, $20\MeV$.
For reference, the $^4$He yield in the absence of a decaying
tau neutrino is $Y_P=0.2228$ (2 massless neutrinos) and
$0.2371$ (3 massless neutrinos).

\bigskip
\noindent{\bf Figure 9:} $^4$He yield as a function of
tau-neutrino lifetime for the $\nu_\tau \rightarrow
\nu_e$ + EM decay mode, $\eta_{0} = 3\times 10^{-10}$, and
Dirac masses of 1, 5, 10, $20\MeV$.
For reference, the $^4$He yield in the absence of a decaying
tau neutrino is $Y_P=0.2228$ (2 massless neutrinos) and
$0.2371$ (3 massless neutrinos).

\bigskip
\noindent{\bf Figure 10:}  $^4$He yield as a function of
lifetime for the $\nu_e + \phi$ decay mode, $\eta_0 = 3\times 10^{-10}$,
$rm = 3.5$, and masses of 5, 10, 20, $30\MeV$.
For reference, the $^4$He yield in the absence of a decaying
tau neutrino is $Y_P=0.2228$ (2 massless neutrinos) and
$0.2371$ (3 massless neutrinos).

\bigskip
\noindent{\bf Figure 11:}  Excluded regions of the $rm - \tau$
plane for the four different decay modes and a $20\MeV$ mass particle.
The limits for the first two decay modes are strictly
independent of mass; for the last two decay
modes they should be relatively insensitive to mass
for $\tau\ga 100\sec$ (and actually more stringent
than those shown here for $m>20\MeV$).  Excluded regions are above the curves.

\bigskip
\noindent{\bf Figure 12:}  Regions of the tau-neutrino mass-lifetime
plane excluded by laboratory experiments and astrophysical
arguments.  The excluded regions are to the side of the
curve on which the label appears.  The dashed curve
summarizes a host of different laboratory limits to
$|U_{e\tau}|^2$, translated to a model-dependent bound
to $\tau_\nu (\nu_\tau \rightarrow \nu_e e^\pm )$, cf.
Eq.~(\ref{UET}).   SNL denotes the supernova-light
constraint which extends to lifetimes shorter than those
shown here.

\end{document}